\documentclass[fleqn,usenatbib]{mnras}

\usepackage[T1]{fontenc}
\usepackage{ae,aecompl}

\usepackage{graphicx}	
\usepackage{amsmath}	
\usepackage{amssymb}	
\usepackage[shortlabels]{enumitem}
\usepackage{lineno}

\usepackage{newtxtext,newtxmath}

\title[A Revised SN~Ia Surface]{A Revised SALT2 Surface for Fitting Type Ia Supernova Light Curves}

\author[G. Taylor et al.]{
G. Taylor$^{1}$\thanks{E-mail: gtaylor@mso.anu.edu.au or SALT2-2021@outlook.com},
C. Lidman$^{1}$,
B. E. Tucker$^{1, 2, 3}$,
D. Brout$^{4, 5}$,
S. R. Hinton$^{6}$,
R. Kessler$^{7, 8}$
\\
$^{1}$Research School of Astronomy and Astrophysics, Australian National University, Canberra, Australia\\
$^{2}$National Centre for the Public Awareness of Science, the Australian National University, Canberra, Australia\\
$^{3}$The ARC Centre of Excellence for All-Sky Astrophysics in 3 Dimensions (ASTRO 3D), Australia\\
$^{4}$Department of Physics and Astronomy, University of Pennsylvania, Philadelphia, PA 19104, USA\\
$^{5}$NASA Einstein Fellow\\
$^{6}$School of Mathematics and Physics, University of Queensland, Brisbane, Australia\\
$^{7}$Department of Astronomy and Astrophysics, University
of Chicago, Chicago, IL 60637, USA\\
$^{8}$Kavli Institute for Cosmological Physics, University of
Chicago, Chicago, IL 60637, USA
}

\date{Accepted XXX. Received YYY; in original form ZZZ}

\pubyear{2021}

\begin{document}
\label{firstpage}
\pagerange{\pageref{firstpage}--\pageref{lastpage}}
\maketitle

\begin{abstract}

We present a revised SALT2 surface (`SALT2-2021') for fitting the light curves of Type Ia supernovae (SNe~Ia), which incorporates new measurements of zero-point calibration offsets and Milky Way extinction. The most notable change in the new surface occurs in the UV region. This new surface alters the distance measurements of SNe~Ia, which can be used to investigate the nature of dark energy by probing the expansion history of the Universe. Using the revised SALT2 surface on public data from the first three years of the Dark Energy Survey Supernova Program (combined with an external low-$z$ SNe Ia sample) and combining with cosmic microwave background constraints, we find a change in the dark energy equation of state parameter, $\Delta w = 0.015 \pm 0.004$. This result highlights the continued importance of controlling and reducing systematic uncertainties, particularly with the next generation of supernova analyses aiming to improve constraints on dark energy properties.

\end{abstract}




\section{Introduction}

Type Ia supernovae (SNe~Ia) are important cosmological probes that are used as distance estimators to help constrain the nature of dark energy (see e.g. \citealt{howell_11} for a review). SNe~Ia are standardisable candles, displaying an empirical relationship between their peak luminosity, and their light curve width (or `stretch') and colour \citep{pskovskii, phil93, riess96, tripp98}. This two-parameter luminosity correction is the most common technique for standardising a set of SNe~Ia. By fitting a light curve model to time-series photometry of SNe~Ia we recover the SN parameters (amplitude, stretch, and colour), and infer their distances. The accuracy at which light curves can be modelled directly impacts these distances, and moreover, cosmological constraints. 

SALT2 (Spectral Adaptive Light Curve Template, \citealt{salt2}) is an empirical spectro-photometric model that is used in most modern SN~Ia analyses, including the SuperNova Legacy Survey 3yr analysis (SNLS, \citealt{conley11}), the Joint Light Curve Analysis (JLA, \citealt{JLA}), the Pantheon analysis \citep{panstarrs18}, the Pan-STARRS1 analysis (PS1, \citealt{Jones_2018}), and the Dark Energy Survey 3yr analysis (DES-SN3YR, \citealt{Abbott_2019}). The SALT2 model consists of two phase-dependent components and
a phase-independent colour law, an instance of which called the SALT2 surface. The SALT2 surface is determined through a training process. The most recent surface is SALT2.4 (hereafter referred to as SALT2-2014), which was produced for JLA, and is publicly available.\footnote{\url{http://supernovae.in2p3.fr/salt/doku.php}} Since 2014 there has been a consensus in the literature for using a revised Galactic dust extinction map, and there have also been improvements in photometric calibrations of some SNe Ia. Dust extinction and calibration are both key inputs to the SALT2 model that will affect the resulting SALT2 surface.

The amount of extinction caused by dust in our Milky Way has been revised by \citet{schlafly2010} and \citet{schlafink}. They find an over-prediction of reddening in the \citet{schlegel98} dust maps, and therefore recommend a 14\% re-calibration of Milky Way extinction as a correction. Changes in the amount of reddening caused by Milky Way dust affect the colour of SNe~Ia, which is an important parameter when fitting SN~Ia distances. This re-calibration was not applied to the SNe~Ia used to train SALT2-2014.

Using the PanSTARRS1 calibration covering 3/4 of the sky, the flux zero-points of some of the SNe~Ia used in the training of SALT2-2014 have been updated by an average of 10~mmag \citep{supercal}. The systematic uncertainty in cosmological parameters arising from our imperfect knowledge of photometric systems and their zero-points is significant. In the JLA, Pantheon, and DES-SN3YR samples, this uncertainty contributes between a quarter and two-thirds of the total systematic uncertainty \citep{JLA, panstarrs18, des3yr_systematics, Jones_2019}.

These improvements in Galactic extinction and photometric calibration are particularly important because statistical and systematic uncertainties are comparable in SN~Ia cosmology analyses \citep{suzuki12, scolnic14}, and the current SALT2 surface has been identified as a limiting factor. For example, the SALT2 model uncertainty contributes $\sigma_w^{sys}$ = 0.009 in \citet{des3yr_systematics}, 0.008 in \citet{Jones_2018} and 0.023 in \citet{Jones_2019}. Current SN~Ia cosmology analyses and future surveys therefore require an improved SALT2 surface to make the most of their high quality data and improve measurements of dark energy properties. Future surveys that focus on producing a large SN sample from a single instrument can also benefit from an updated SALT2 surface, as they will see the effects of the new surface.

Here, we update the SALT2 surface using recent improvements in calibration and Galactic extinction to produce a new surface, SALT2-2021\footnote{Available at 10.5281/zenodo.4646495}. When using SALT2-2021, the revised Galactic extinction map and Supercal calibration adjustments should be applied to the data used for light curve fitting. We include an overview of the SALT2 model, including the training and fitting processes ($\S$\ref{training}). We then describe the specific changes made to produce the SALT2-2021 surface ($\S$\ref{changes}). We evaluate the impact of SALT2-2021 on cosmological distance measurements using the DES-SN3YR sample and simulations ($\S$\ref{results_big}), and discuss these results ($\S$\ref{discussion}).

\section{SALT2}
\label{training}

\subsection{Model Description}

SALT2 is an empirical SNe~Ia model \citep{salt, salt2}. The model describes the spectro-photometric time evolution of a SN~Ia, and was built using both spectroscopic and photometric data from a large set of nearby ($z~\lesssim~0.1$) and distant ($0.2~\lesssim~z~\lesssim~0.8$) SNe~Ia. 

In SALT2, the spectral flux density for each SN~Ia at a particular phase ($p$) and wavelength ($\lambda$) is given as
\begin{linenomath*}
  \begin{align} 
    \label{flux1}
    \begin{split}
    f_{\lambda} = x_{0} &\times\left[M_{0}(p, \lambda)+x_{1} M_{1}(p, \lambda)+\ldots\right] \\
    &\times \exp [c CL(\lambda)],
    \end{split}
  \end{align}
\end{linenomath*}
where amplitude ($x_0$), stretch ($x_1$), and colour ($c$) are parameters determined from a light curve fit for each SN~Ia; $M_0$ and $M_1$ are phase-dependent components of the SALT2 model; and $CL$ is the phase-independent SALT2 colour law. Specifically, $M_0$ describes the mean spectral energy distribution (SED) of a SN~Ia, while $M_1$ describes the first-order deviation around this SED.

The SALT2 model does not fully describe all the observed SN~Ia variability --- the remaining variability ($\sim$~0.1~magnitudes) is often labelled as `intrinsic scatter'. Scatter in the SN population itself can have coherent and chromatic contributions. To simulate distance biases, commonly used intrinsic scatter models are from \citet{Kessler_2013} --- these models are based on \citet{g10_scatter} ($\sim$~75\% coherent contributions, $\sim$~25\% chromatic contributions) and \citet{c11_scatter} ($\sim$~25\% coherent contributions, $\sim$~75\% chromatic contributions).  Currently, there is no strong evidence favouring one model over the other.

To understand the origins of the scatter and to further improve SNe Ia as standard candles, a number of studies have looked for and found evidence for relationships between the properties of the SN and its host galaxy. Relationships have been found between the luminosity of a SN~Ia and its host galaxy metallicity, morphological type, and mass \citep{hamuy00, wang06, kelly2010, sullivan_host, smith2020cosmology}. \citet{scolnic_siblings} use a different approach and consider the properties of SN Ia `siblings' (i.e. SNe Ia that share the same parent galaxies) and find at least 50\% of the intrinsic scatter of SNe Ia distance modulus residuals does not originate from common host properties, albeit using a sample of only 8 pairs of SN Ia siblings. A similar study of two SNe Ia in the same galaxy (NGC 3972) finds significant luminosity differences between SNe Ia with similar light curve shapes and colours, reaffirming that the popular two-component parameterisation (as in SALT2) does not fully describe SNe Ia \citep{foley_twins} without intrinsic scatter.

An additional complication from host galaxies is their dust extinction (reddening) contribution. As it is difficult to disentangle the effects of dust and the intrinsic relation between SN colour and brightness \citep{nobili03}, contributions from both are implicitly included in the colour ($c$) parameter and colour law ($CL$) of SALT2. 

Alternatives to SALT2 have different approaches to light curve modelling. We briefly outline some differences between SALT2 and other common models. More in-depth comparisons between light curve models can be found in the literature (e.g. \citealt{compare, kessler09}). We focus on SALT2 as it is publicly available and is widely used in SN analyses (e.g. JLA, Pantheon, DES).

MLCS2K2 \citep{mlcs} and SNooPy \citep{snoopy} attempt to explicitly separate reddening by dust from phase-dependent intrinsic colour, whereas SALT2 combines both effects into one colour parameter. SiFTO \citep{sifto} performs similarly to SALT2, but differs with respect to SALT2 in the way colour and stretch are treated (see \citealt{g10_scatter} for details). SUGAR \citep{sugar} has more parameters than SALT2, and derives a colour law that is consistent with Milky Way extinction (whereas the SALT2 colour law differs significantly from the reddening from dust in the Milky Way).

\subsection{Training}
\label{trainingprocess}

\begin{figure}
  \begin{center}
   \includegraphics[width=\columnwidth]{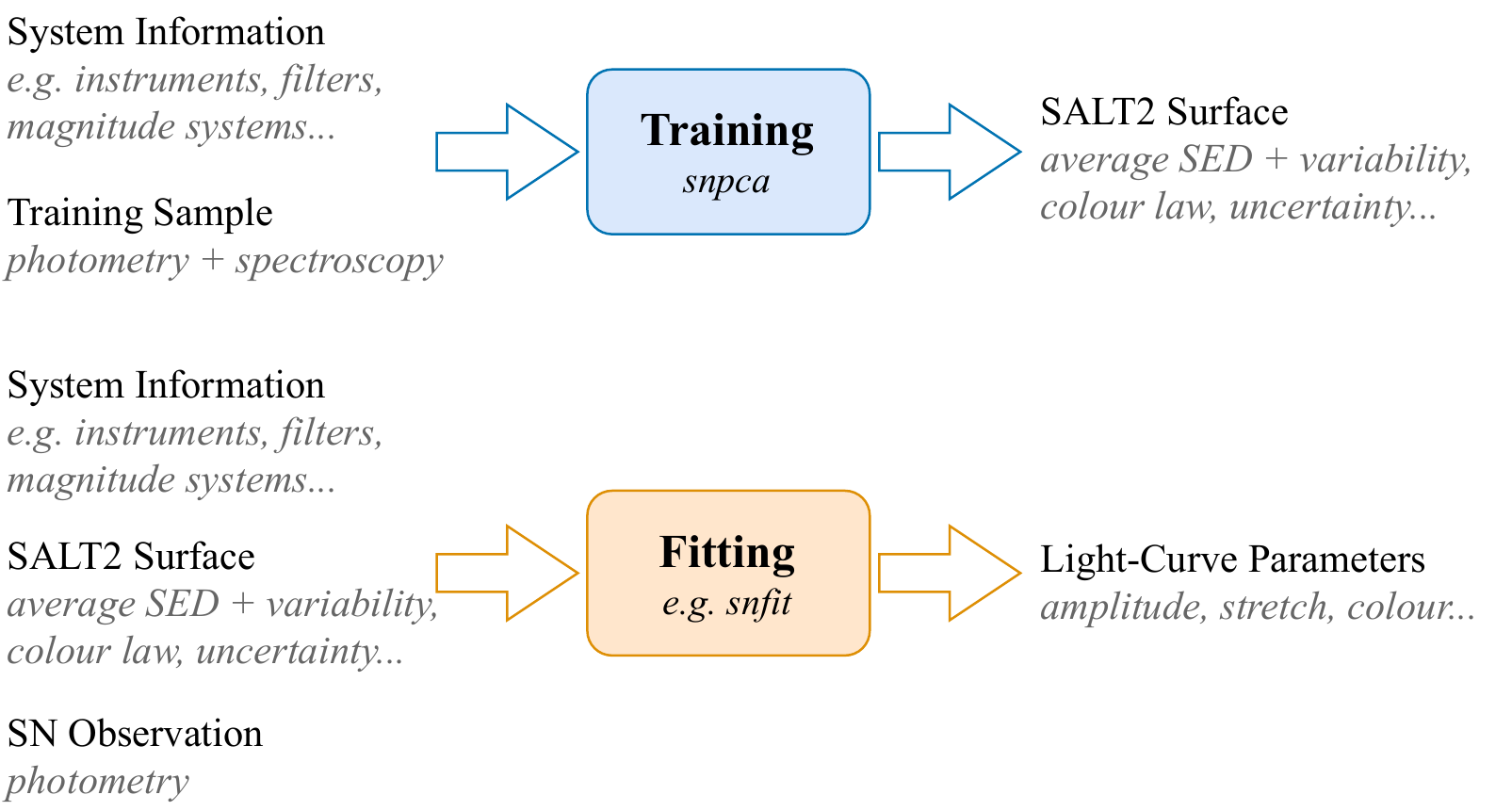}
  \end{center}
  \caption{An illustration of the inputs and outputs of the SALT2 training and fitting processes. The training of the global SALT2 model produces a SALT2 `surface', which is used as an input for light curve fitting. The recovered light curve parameters are therefore dependent on the SALT2 surface used. The surface needs to be recomputed when the inputs, e.g. magnitude systems, are revised.}
  \label{Schematic}
\end{figure}

The progenitor and explosion mechanisms leading to SNe~Ia are areas of active research \citep{Maoz_2014}, and theoretical models producing SN~Ia spectra are not yet good enough to be used for correcting SN~Ia brightness. For these reasons, models are empirically determined from a training sample of real SN observations. The original SALT model \citep{salt} was developed into the current SALT2 model \citep{salt2}. Two widely used SALT2 surfaces have been released: SALT2.2, an SNLS release \citep{g10_scatter}; and SALT2.4, released as part of the joint light curve analysis of SNLS and SDSS-II SNe~Ia (\citealt{JLA}, here referred to as SALT2-2014).

SALT2 is trained on a sample of photometric and spectroscopic observations of both nearby and distant SNe~Ia. Including spectral data improves modelling of spectral features that are hard to access using photometry alone, while including higher-redshift SNe reduces the dependence on the nearby SNe sample and provides better constraints on the rest-frame ultra-violet regions of the spectrum \citep{compare}. This training sample, along with information on the instruments and magnitude systems, is passed to the \verb|snpca|\footnote{We use v2.3.22, which was kindly provided to us by Marc Betoule.} training program, as shown in Figure~\ref{Schematic}. The output of \verb|snpca| is a SALT2 surface, i.e. an iteration of the SALT2 model. For a detailed description of the training process, see \citet{mosher14}.

\subsection{Fitting Light Curves, Distances, and Cosmology}

A SALT2 surface is passed to a light curve fitting program to fit light curves to photometric time-series observations of SNe. Available fitting programs include \verb|snfit|\footnote{\url{http://supernovae.in2p3.fr/salt/doku.php}}, \verb|SNANA|\footnote{\url{http://snana.uchicago.edu/}}, and \verb|SNCosmo|\footnote{\url{https://sncosmo.readthedocs.io/}}. This fit produces the most likely light curve parameters $x_0$, $x_1$ and $c$ from Equation~\ref{flux1}, together with their covariances. From here, distance moduli ($\mu$) are obtained using the Tripp equation \citep{tripp98}:
\begin{linenomath*}
  \begin{align}
  \label{trippeq}
    \mu = m_B - M + \alpha x_1 - \beta c,
  \end{align}
\end{linenomath*}
where $m_B = -2.5\log(x_0)$, $M$ represents the absolute magnitude of a SN~Ia with $c=x_1=0$, and $\alpha$ and $\beta$ are nuisance parameters representing the slopes of the stretch-luminosity and colour-luminosity relations. The nuisance parameters ($\alpha, \beta, M$) can be determined together with cosmological parameters (e.g. \citealt{salt, JLA}), or in a separate step before the cosmological parameters are fitted (e.g. \verb|SALT2mu|, \citealt{salt2mu}; \verb|BBC|, \citealt{Kessler_2017}). 

Some modifications to the Tripp equation (in the form of an additional `mass-step' term) have been used to account for the observed correlation between SN brightness and host-galaxy properties, e.g. \citet{kelly2010, sullivan_host}. Following \citet{des3yr_systematics}, we include a mass-step correction ($\delta \mu_{host}$). We also include a 5-D bias-correction term from \verb|BBC| ($\delta \mu_{bias}$), which uses detailed simulations to predict biases from selection effects. The modified version of the Tripp equation used in this work is therefore
\begin{linenomath*}
  \begin{align}
  \label{modified_trippeq}
    \mu = m_B - M + \alpha x_1 - \beta c - \delta \mu_{\rm host} - \delta \mu_{\rm bias}.
  \end{align}
\end{linenomath*}

\section{Improvements in the New Surface (SALT2-2021)}
\label{changes}

We have produced a new surface denoted SALT2-2021 that implements two important updates to the model inputs. The underlying SALT2 model and code used to produce this surface are identical to that used for SALT2-2014, with the exception that we sample the resulting surface more finely in wavelength.

\subsection{Training Sample}
\label{trainsample}

\begin{table*}
  \renewcommand{\arraystretch}{1.5}
  \centering
  \caption{SNe~Ia used in SALT2 training sample. 2 SNe were observed by SDSS, CSP, and CfA3, and 4 SNe were observed by both SDSS and CfA3 - giving a total of 420 unique SNe Ia in the training sample.}
  \label{tab:training sample}
  \begin{tabular}{ccc}
    \hline
		Survey & Number of SNe & Redshift range\\ 
		\hline
		SDSS-II \citep{sdssII} & 203 & 0.1~$\lesssim~z~\lesssim$~0.3\\
		SNLS \citep{conley11, sullivan11} & 113 & 0.2~$\lesssim~z~\lesssim$~0.9\\
		Calan/Tololo \citep{hamuy_96} & 5 & $z~\lesssim$~0.1\\
		CSP \citep{csp} & 2 & $z~\lesssim$~0.1\\
		CfA1 \citep{riess_99} & 8 & $z~\lesssim$~0.1\\
		CfA2 \citep{jha_06} & 14 & $z~\lesssim$~0.1\\
		CfA3 \citep{Hicken_2009_cfa3} & 58 & $z~\lesssim$~0.1\\
		Historical low-$z$ (\citet{JLA} and references within) & 25 & $z~\lesssim$~0.1\\
		\hline
  \end{tabular}
\end{table*}

\begin{figure}
  \begin{center}
   \includegraphics[width=\columnwidth]{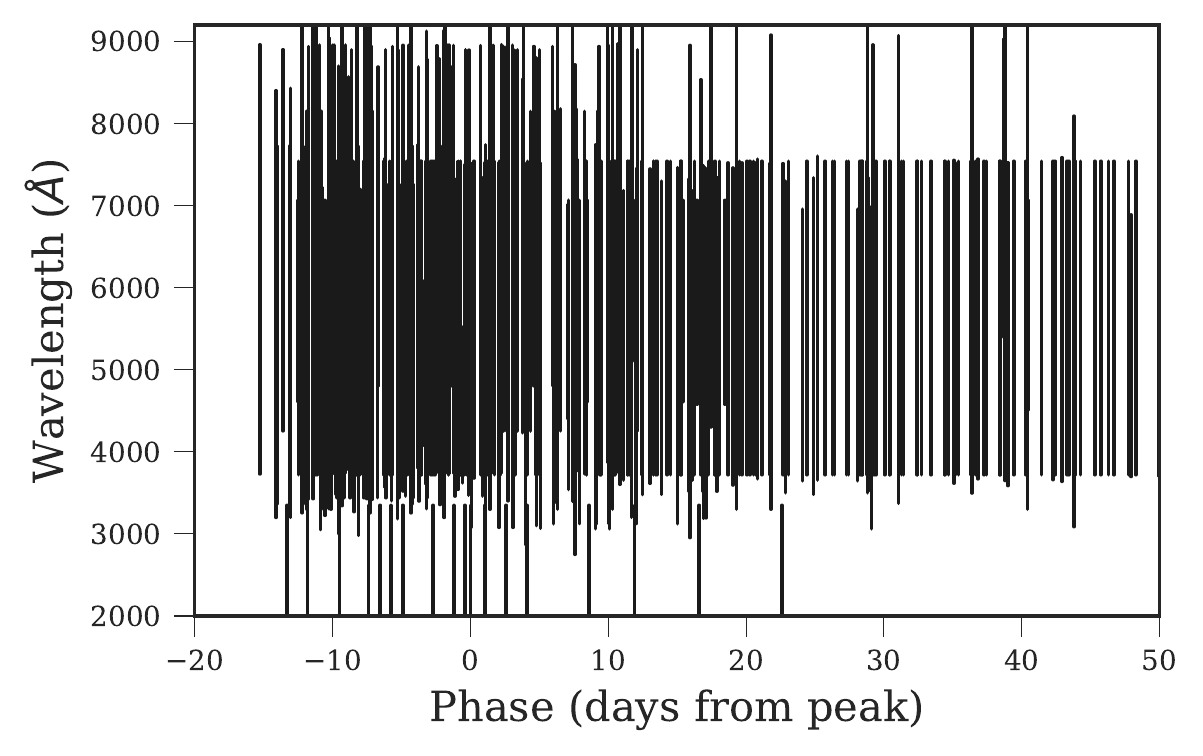}
  \end{center}
  \caption{Wavelength coverage of the spectra used in the training sample, plotted as a function of phase.}
  \label{spec_coverage}
\end{figure}

To compare our results with the previous SALT2-2014 surface \citep{JLA}, we use the same training sample of SNe. This sample contains photometry and spectroscopy from 420 SNe~Ia from various surveys, listed in Table~\ref{tab:training sample}. All 420 SNe Ia in the training sample have photometric data, but only 83 of those have spectral data. The phase and wavelength coverage of this spectral data is shown in Figure~\ref{spec_coverage}. For more details on this training sample, see \citet{JLA}.

Altering the set of SNe used in the training sample is beyond the scope of this work, which focuses primarily on the effects of modifying the Milky Way dust extinction and photometric zero-points on the SALT2 surface. In future work, we will explore how the SALT2 surface changes when we add more recent SN Ia data.

\subsection{Re-Scaled Milky Way Extinction}
\label{mwebv}

A reddening \textit{law} predicts the amount of absorption versus wavelength, relative to the absorption at a reference wavelength (typically the effective wavelength of the Bessell B band). A \textit{dust map} predicts the reddening for the reference wavelength at any sky location. Together, a dust map and reddening law predict the amount of absorption at any wavelength, and any sky location.

When fitting a light curve, SALT2 uses the \citet{ccmdust} reddening law to correct SNe Ia photometry for Milky Way extinction, with the colour excess $E(B-V)$ calculated from the \citet{schlegel98} dust maps.\footnote{The SALT2 colour law derived during the training process includes contributions from the intrinsic colour of an SN, as well as any host galaxy extinction.} A 14\% re-calibration of the \citet{schlegel98} dust map is prescribed in \citet{schlafly2010} and \citet{schlafink}. Until now, this re-calibration had not been applied to the SNe used to train SALT2. We implement this re-calibration by adjusting the $E(B-V)$ of all the SNe in the SALT2 training sample by a factor of 0.86, such that 
\begin{linenomath*}
  \begin{align}
  \label{ebveq}
    E(B-V)_{\rm new} = 0.86 \times E(B-V)_{\rm old}.
  \end{align}
\end{linenomath*}

The \citet{schlafink} dust maps are the most commonly used dust maps for supernova analyses (e.g. \citealt{Burns_2014, supercal, asassn_2013, Foley_2016, Pierel_2018, panstarrs18, Dimitriadis_2018, Kessler_2019_sims, khetan2020new, BayeSN}). Other dusts maps, for example the dust map derived using data from the Planck satellite \citep{planckdust}, differ slightly from the the map used here. We evaluate the impact of the dust maps by training a version of SALT2-2021 with \citet{planckdust} dust maps (instead of \citealt{schlafink}) and re-running the analysis in $\S$~\ref{results_big}. We find $\Delta w = 0.01$ compared to the results using the \citet{schlafink} map. This is a slightly smaller than the shift in $w$ from adopting the revised zero points and the re-calibration of the \citet{schlafink} dust maps.

\subsection{Re-calibrated Photometry}
\label{zps}

The SALT2-2014 surface is trained on SNe from a number of surveys ($\S$\ref{trainsample}) with little or no overlapping sky area, and thus the calibration relies on a standard star system. The approach used by these surveys to calibrate the flux scale is not homogeneous. This introduces systematic errors that can be difficult to quantify, and which propagate through to redshift-dependent distance biases (resulting in a $w$ bias).

`Supercal' \citep{supercal} improves the relative calibration by using tertiary standard stars in regions of survey overlap to determine the zero-point offset between each system and the Pan-STARRS1 (PS1) system, which has a relative calibration uncertainty of $<$5~mmag \citep{ps1.5, ps1} --- thus effectively re-calibrating each system. The effect of this re-calibration was to shift $w$ by 0.026 for a combined sample that includes PS1, SNLS, and SDSS \citep{supercal}.

Until now, this re-calibration has only been applied to samples \textit{fitted} with the SALT2 model; however, the SALT2-2014 surface has been \textit{trained} without Supercal. We adjust the SALT2 training data to include the super-calibrated zero-points and retrain SALT2, to ensure consistent photometric calibration for future SN analyses. This re-calibration affected 390 out of 420 SNe~Ia in the training sample --- the unaffected 30 SNe are from surveys that were not re-calibrated by Supercal. The adopted zero-point offsets are in Table~5 of \citet{panstarrs18}.

\subsection{Calibration Uncertainties}

In addition to the SALT2-2021 surface, we provide a suite of perturbed surfaces, where each perturbed surface has been trained with a shift applied to one zero point (0.01 mag) or 1 filter transmission curve (1 nm). These perturbed surfaces span all instruments and filter bands used in the training, and can be
used to account for correlated systematic uncertainties in the training and light curve fitting (e.g., see $\S$~5.4 in B14). The perturbed surfaces are available online.\footnote{10.5281/zenodo.4646495}

\subsection{Comparing SALT2-2021 and SALT2-2014}
\label{salt2-2021}

\begin{figure}
  \begin{center}
   \includegraphics[width=\columnwidth]{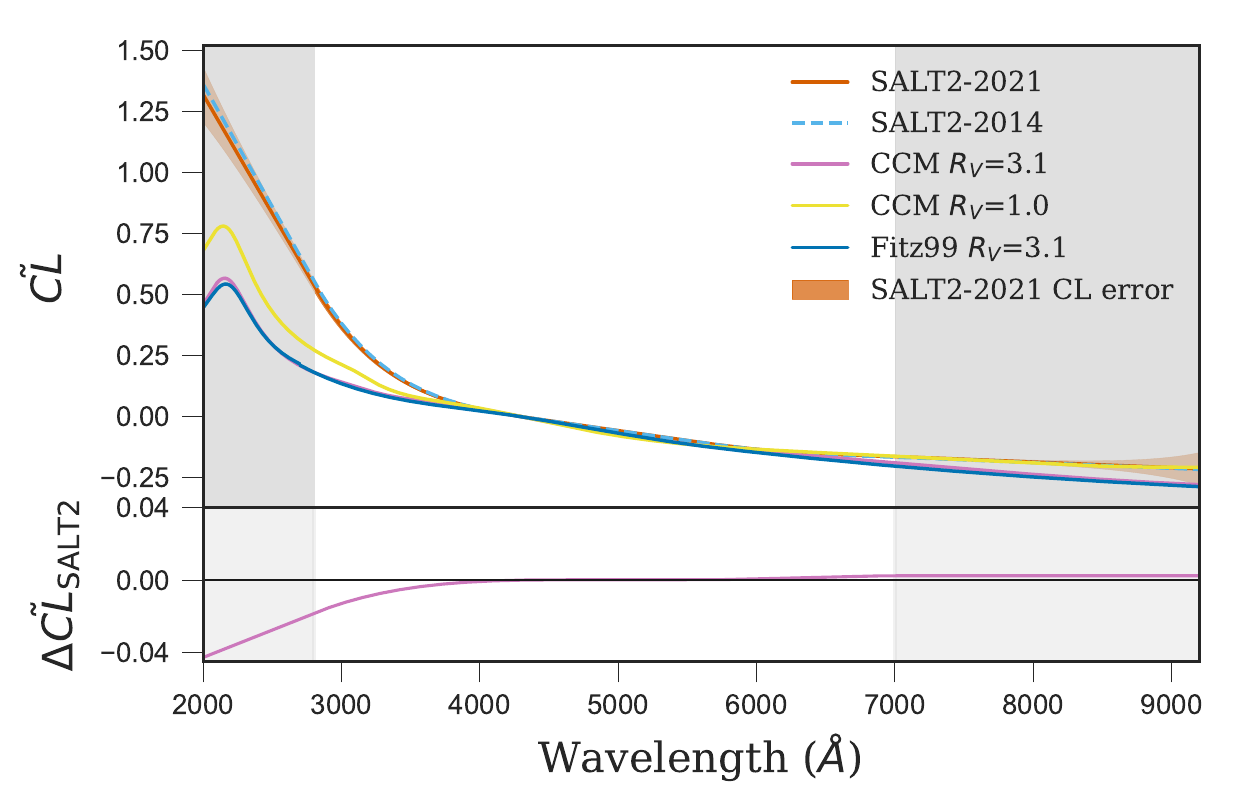}
  \end{center}
  \caption{The derived colour law ($CL$) for the old \textit{(cyan, dashed)} and new \textit{(orange, solid)} SALT2 surfaces, for a $c=0.1$ SN~Ia, where $\tilde{CL} \equiv - c CL(\lambda)$ (for the SALT2 colour laws). The white region is the wavelength range over which the SALT2 colour law is fit, and the grey shaded regions represent the wavelength ranges in which the SALT2 colour law is linearly extrapolated from the end points. Also plotted are common extinction laws from \citealt{ccmdust} (CCM) and \citealt{fitz99} (Fitz99), for $E(B-V)~=~0.1$, where $\tilde{CL} \equiv A(\lambda)-A(B)$ for the extinction laws. Differences between the SALT2-2021 and SALT2-2014 surfaces are shown in the lower panel.}
  \label{CL}
\end{figure}

\begin{figure}
  \begin{center}
   \includegraphics[width=\columnwidth]{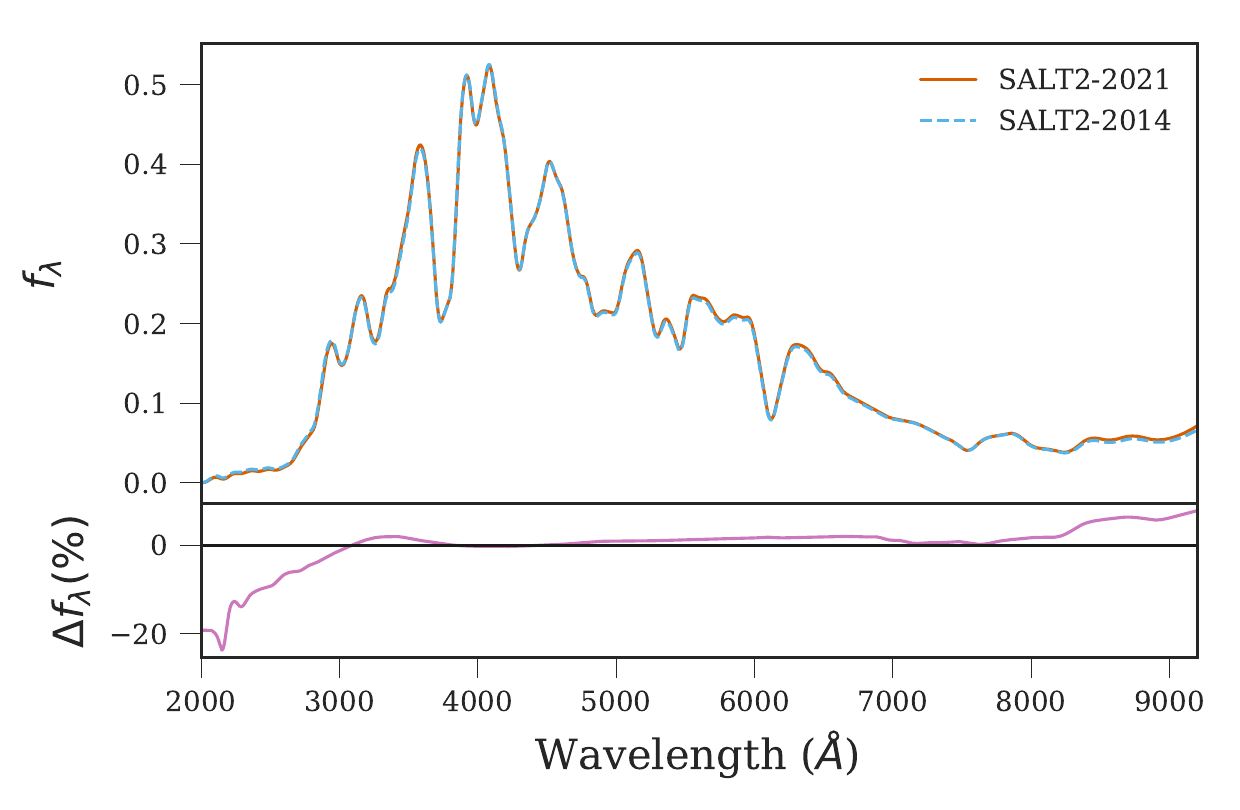}
  \end{center}
  \caption{The trained spectral energy distribution of SNe~Ia ($c$=0, $x_1$=0) for the old \textit{(cyan, dashed)} and new \textit{(orange, solid)} SALT2 surfaces. The lower panel shows the difference in the average spectral energy distribution between the old and new SALT2 surfaces at each wavelength, where $\Delta f_\lambda (\%) = 100 \times \frac{f_{\lambda 2021} - f_{\lambda 2014}}{f_{\lambda 2021}}$.}
  \label{SED}
\end{figure}

\begin{figure}
  \begin{center}
   \includegraphics[width=\columnwidth]{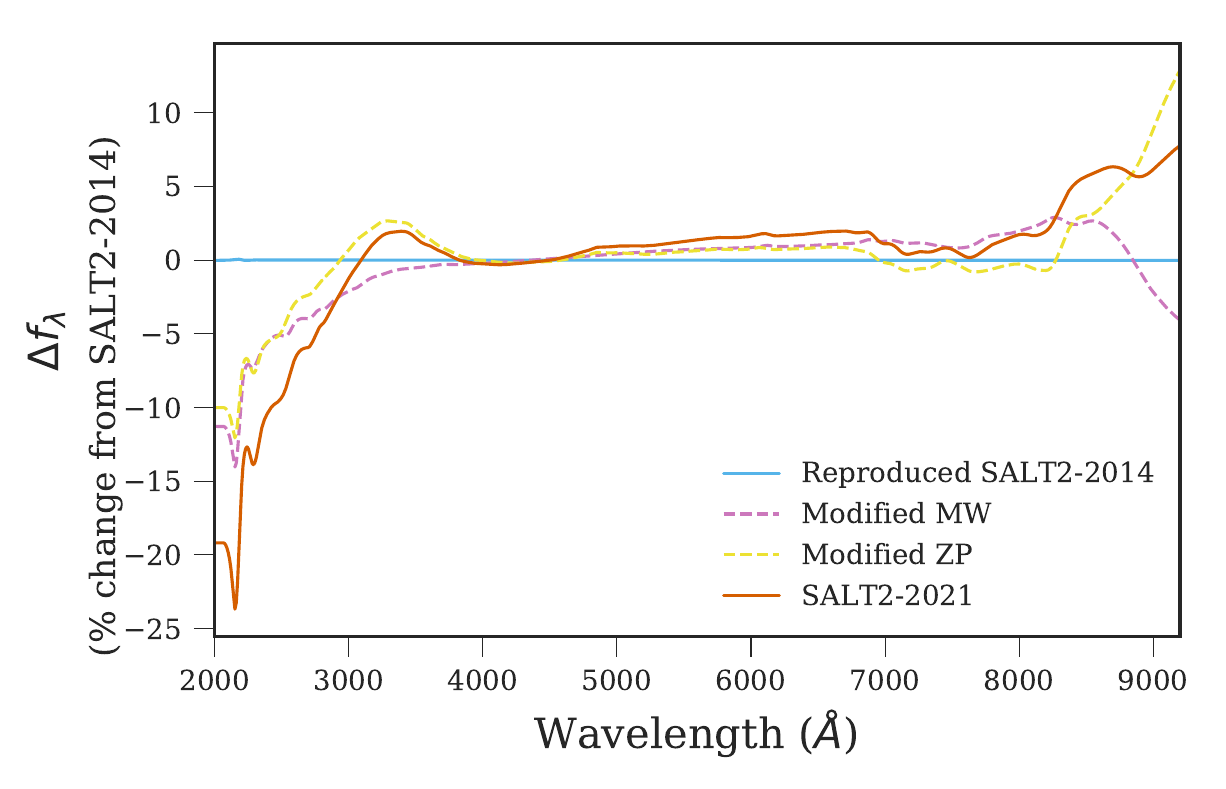}
  \end{center}
  \caption{The difference in the spectral energy distribution between the SALT2-2014 surface, and: a reproduced version of SALT2-2014 for process verification (\textit{cyan}, Appendix~\ref{verification}); a retrained surface with only re-scaled Milky Way extinction applied (\textit{magenta}, $\S$\ref{mwebv}); a retrained surface with only re-calibrated zero-points applied (\textit{yellow}, $\S$\ref{zps}); and the final SALT2-2021 surface with Milky Way and zero-point adjustments applied (\textit{orange}). We are able to reproduce the original SALT2-2014 surface to within 0.1\%, whereas the new SALT2-2021 surface has fractional changes of up to 20\% relative to SALT2-2014 (albeit at the low-flux ends of the spectrum).}
  \label{SEDdiffs}
\end{figure}

The SALT2-2021 colour law for a $c=0.1$ SN~Ia, plotted in Figure~\ref{CL}, agrees with the SALT2-2014 colour law for wavelengths above $\sim$~4000~\AA. Below $\sim$~4000~\AA, there is some deviation. The SALT2 colour law differs significantly from the \citet{ccmdust} and \citet{fitz99} colour laws (also plotted in Figure~\ref{CL}), again, mainly in the region below $\sim$~4000~\AA. This difference is long-known and shows that there is an intrinsic component that has more impact in the UV than dust.

The average spectral energy distribution of the revised SALT2-2021 surface, plotted in Figure~\ref{SED}, is in good agreement with the SALT2-2014 surface between $\sim$~3000-8000~\AA. There is a difference of a few percent above $\sim$~8000~\AA, and much larger difference below $\sim$~3000~\AA, reaching 20\% in the UV part of the spectral energy distribution. This region ($\sim$~2000~\AA) is poorly sampled in the training set for SALT2 (as shown in Figure~\ref{spec_coverage}) and therefore poorly constrained. These effects also correspond to the linearly-extrapolated region of the SALT2 colour law (the shaded region of Figure~\ref{CL}), where the changes are greatest. 

Figure~\ref{SEDdiffs} shows the differences between the published SALT2-2014 surface, and various iterations of the new surface (trained with only Milky Way extinction modified, with only the zero points modified, and with both modified). We are able to reproduce the original SALT2-2014 surface to within 0.1\%, giving us confidence in our training procedures (see Appendix~\ref{verification} for details). The differences in the final SALT2-2021 spectral energy distribution are the sum of differences due to modifying the Milky Way extinction and modifying the photometric zero points.

\section{First Analysis with SALT2-2021}
\label{results_big}

\subsection{The DES-SN3YR Sample}
\label{des_sample}

The Dark Energy Survey (DES, \citealt{des_summary}) is a multi-probe investigation into the nature of dark energy, using DECam \citep{decam, Morganson_2018} on the 4m Blanco telescope at the Cerro Tololo Inter-American Observatory (CTIO). DES operated from 2013-2019 and collected data for a 5000~deg$^2$ wide-field survey, as well as a 27~deg$^2$ dedicated Dark Energy Survey Supernova Program (DES-SN, \citealt{dessnsims}). Over 30,000 transients were discovered with DES-SN, and light curves were measured in $griz$ bands over a roughly weekly cadence \citep{des_diffimg}. A subset of these transients have been identified as SNe~Ia, via either photometric ($\sim$~3000 SNe from 0.01~$\leq~z~\leq$~1.2) or spectroscopic ($\sim$~500 SNe from 0.017~$\leq~z~\leq$~0.9) classification \citep{desspec}.

The sample we analyse in this work is from the published set of 207 spectroscopically confirmed SNe~Ia from the first three years of the Dark Energy Survey, combined with a selection of 122 low redshift SNe~Ia (DES-SN3YR, \citealt{des3yr_systematics}), released as part of the Dark Energy Survey's Data Release I \citep{des_dr1}.\footnote{Available from \url{https://des.ncsa.illinois.edu/releases/sn}}. The external low redshift subset includes SNe Ia ($0.01~<~z~<~0.1$) from the Harvard-Smithsonian Center for Astrophysics surveys (CfA3, CfA4; \citealt{Hicken_2009_cfa3}, \citealt{cfa4}) and the Carnegie Supernova Project (CSP; \citealt{csp}, \citealt{csp_2}).

Though we begin with the full sample of 329 SNe from the DES-SN3YR sample, our results are subject to selection requirements (cuts) applied in different stages of the analysis. After these cuts are applied, we are left with 326 SNe using SALT2-2014, and 325 SNe using SALT2-2021 (the discarded SNe are given in Table~\ref{tab:dropped_sn}). The difference in the number of SNe that pass cuts for each surface is due to SN1298893. Fitting SN1298893 with SALT2-2021 changed the recovered PEAKMJD value by $+17$ days, compared with using SALT2-2014. This caused the recovered Trestmin value (time of earliest observation used in fit, relative to PEAKMJD) to change from $-6.6$ to $+11$ days --- the fitting requires $-20 \leq \rm TRESTMIN \leq 0$, so SN1298893 was cut from the SALT2-2021 data set, but not the SALT2-2014 data set. In this case, there is a subtle fit instability originating from an initial PEAKMJD estimate that is more than a week off the final SALT2-2014 value. The SALT2-2014 analysis pushes the fit one way, and the SALT2-2021 analysis pushes the fit in the opposite direction, causing a catastrophic fit. This issue is inherent to the light curve fitting process, not the SALT2 surfaces. Following the DES-SN3YR analysis, a more robust pre-fit PEAKMJD estimate has been developed that fixes this catastrophic fit. For this re-analysis, however, we use the original DES-SN3YR fitting algorithms.

\begin{table*}
\renewcommand\thempfootnote{\arabic{mpfootnote}}
\begin{minipage}{\textwidth}
  \renewcommand{\arraystretch}{1.5}
  \centering
  \caption{Parameters of the DES-SN3YR SNe that did not pass our cuts. Values are from the published DES-SN3YR data.}
  \label{tab:dropped_sn}
  \begin{tabular}{cccccccc}
    \hline
		SN & Sample & $x_1$ & $c$ & $m_B$ & $z$ & Cut at stage & Cut for surface\\ 
		\hline
		1298893 & DES & -2.23698 & -0.0279837 & 20.8162 & 0.1962 & Light curve fitting\footnote{One object failed the light curve fitting requirement $-20 \leq \rm TRESTMIN \leq 0$.} & SALT2-2021\\
		1330031 & DES & 0.022023 & -0.128202 & 18.7782 & 0.104 & Bias corrections\footnote{Three objects with no 5D bias corrections were rejected at the \texttt{BBC} stage. \texttt{BBC} was unable to determine bias correction because of limited MC coverage in this region of $z$, $c$, $x_1$.} & SALT2-2014, SALT2-2021\\
		1308884 & DES & -1.42624 & 0.072173 & 19.0801 & 0.0772 & Bias corrections & SALT2-2014, SALT2-2021\\
		2002G & Low-$z$ & -1.52225 & 0.2214 & 17.3307 & 0.03509 & Bias corrections & SALT2-2014, SALT2-2021\\
		\hline
  \end{tabular}
\end{minipage}
\end{table*}

\subsection{Analysis Overview}

We perform two analyses - one using SALT2-2014, and the other using SALT2-2021 - to understand the impact of our new SALT2 surface on a cosmology analysis. Each analysis follows \citet{des3yr_systematics}, with the general outline as follows: 

\begin{enumerate}
\item We generate detailed simulations \citep{Kessler_2019_sims} for bias corrections.
\item We fit light curves from DES-SN3YR data and simulations.
\item A \verb|BBC| fit applies 5D bias corrections, and fits for $\alpha$, $\beta$, and distance moduli (Equation~\ref{modified_trippeq}) averaged in 20 redshift bins.
\item We create a (statistical-only) covariance matrix for use in cosmology fitting.
\item Using binned distances from \verb|BBC|, we fit a flat $w$CDM cosmology to measure $\Omega_m$ and $w$. We fit these cosmological parameters using the DES-SN3YR data, combined with \citet{planck_cmb} cosmic microwave background measurements. We use the fitted cosmological parameters to calculate $\Delta w$ arising from the change in SALT2 surface, i.e. $\Delta w \equiv w($SALT2-2021$) - w($SALT2-2014$)$.
\end{enumerate}

We perform the simulations, light curve fitting, and bias corrections using \verb|SNANA| \citep{snana09, Kessler_2019_sims}, a publicly-available supernova analysis software package. The cosmology fitting is performed with \verb|CosmoMC| \citep{cosmomc}. We run these programs inside the \verb|Pippin| \citep{Hinton2020} pipeline for SN analyses.

\subsection{Results}
\label{results}

\begin{figure*}
  \begin{center}
   \includegraphics[width=2\columnwidth]{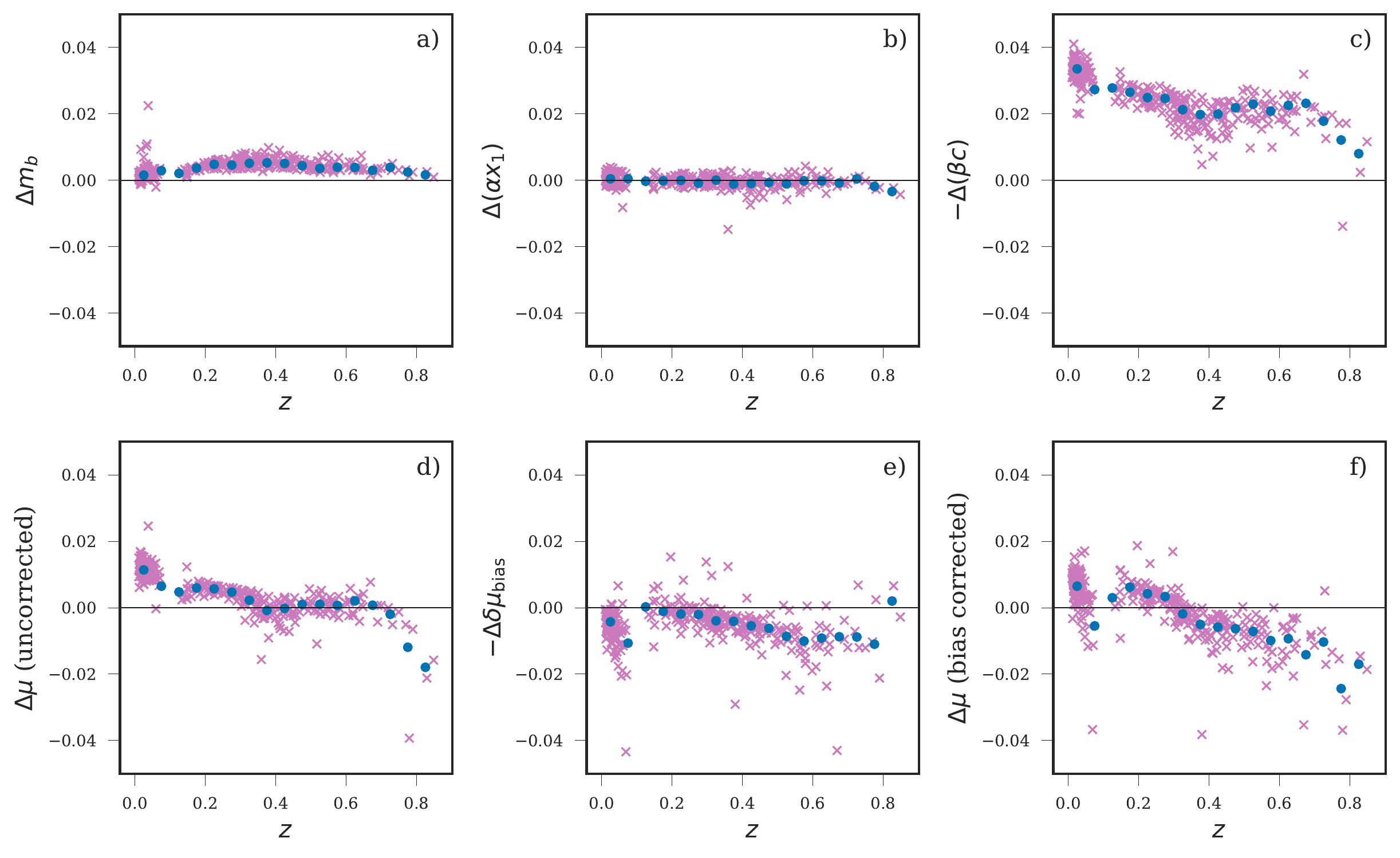}
  \end{center}
  \caption{Differences in fitted parameters between the old and new SALT2 surfaces, plotted versus redshift. Differences are given as e.g. $\Delta \mu = \mu_{2021} - \mu_{2014}$.
  The top row shows the (uncorrected) supernova parameters as they appear in Equation~\ref{trippeq}: $m_B$, $\alpha~\times$~stretch, and $- \beta~\times$~colour.
  Together with $M$ (where $\Delta M_{\rm avg}$ = -0.024, from Table~\ref{tab:alphabeta}, these sum to the (pre-bias corrected) distance modulus.
  The bottom row shows: pre-bias corrected $\Delta \mu$, the change in bias corrections applied to $\mu$, and $\Delta \mu$ after bias corrections. 
  Note that each term is plotted with its sign according to Equation~\ref{trippeq}, so that the true impact on $\mu$ is shown in each panel.
  Each SN in the fitted sample is plotted as a pink cross, with the binned averages plotted as blue circles \textit{(nbins=20)}. One outlier, SN~2007kh, is not displayed.}
  \label{fivepanel}
\end{figure*}

\begin{figure}
  \begin{center}
   \includegraphics[width=\columnwidth]{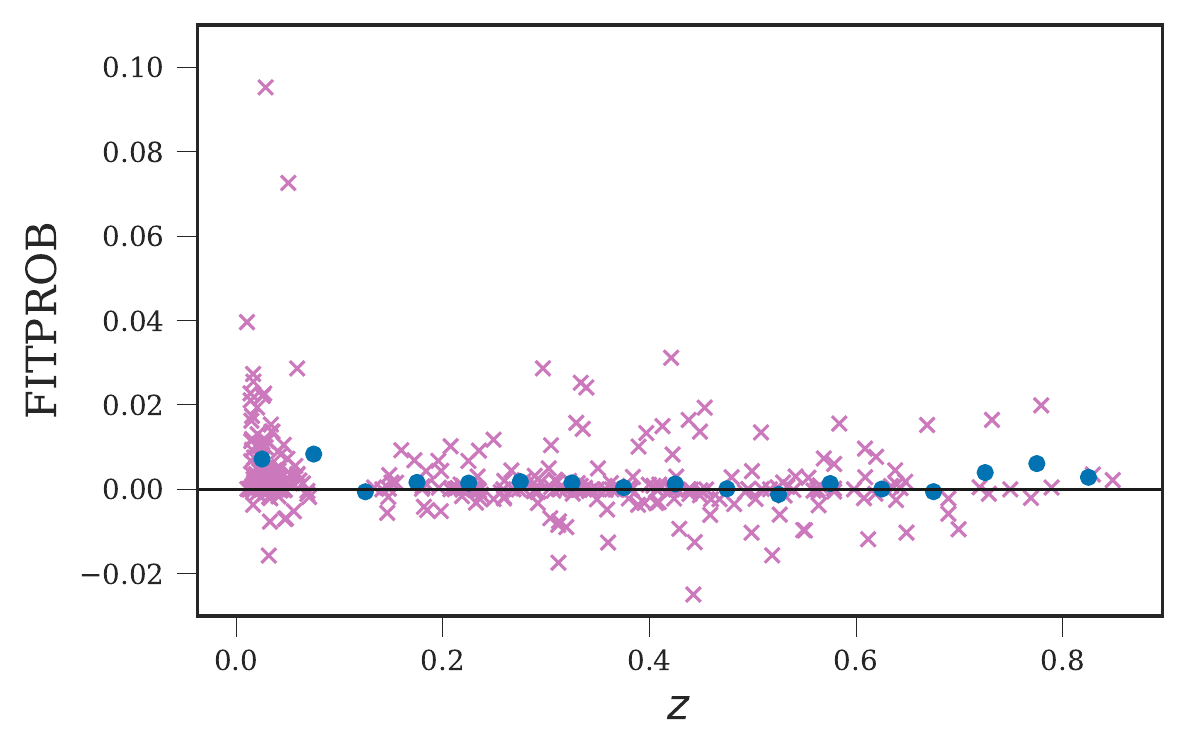}
  \end{center}
  \caption{Differences in the FITPROB parameter between the old and new SALT2 surfaces ($\Delta \rm FITPROB = \rm FITPROB_{2021} - \rm FITPROB_{2014}$), plotted versus redshift. The FITPROB parameter describes the light curve fit probability, from 0 to 1. Each SN in the fitted sample is plotted as a pink cross, with the binned averages plotted as blue circles \textit{(nbins=20)}. One outlier at (0.012,~0.147) is not displayed.}
  \label{fitprob}
\end{figure}

We fit light curves to the DES-SN3YR data and simulations using each surface, and compare the resultant differences in fitted light curve parameters (Equation~\ref{modified_trippeq}) in Figures~\ref{fivepanel}a - \ref{fivepanel}c. At this stage of the analysis, bias corrections have not been applied to the fitted light-curve parameters. We find a clear systematic shift in the distribution of supernova colour represented as the $\beta c$ component. SNe Ia have bluer colours and are brighter when fitted by SALT2-2021, so the change in absolute magnitude ($M_{\rm avg}$) of -0.024 (Table~\ref{tab:alphabeta}) largely compensates for this change in colour. There is a smaller shift in the distribution of stretch represented as the $\alpha x_1$ component, most noticeable (but still minor) at higher redshifts. A small systematic shift in the fitted $m_B$ parameter is also shown. The overall fit probability of the light curves tends to improve (i.e. increase) with SALT2-2021, as shown in Figure~\ref{fitprob} - this effect is most significant in the low-redshift subset.

We compute and apply 5D-bias corrections for distance moduli using \verb|BBC|, with each bias correction simulation generated from the same SALT2 surface used in light-curve fitting. The fitted light-curve parameter distributions for the DES-SN3YR data and the bias correction simulations were compared, and matched well. The recovered nuisance parameters $\alpha$, $\beta$, $\sigma_{\rm int}$ and $M_{\rm avg}$\footnote{From Equation~\ref{modified_trippeq}. M is calculated and applied for each redshift bin; we report $M_{\rm avg}$ for convenience.} are shown in Table~\ref{tab:alphabeta} - these change by $<~1\%$.

\begin{table}
  \centering
  \renewcommand{\arraystretch}{1.5}
  \caption{The recovered nuisance parameters for the DES-SN3YR sample using the old and new surfaces.}
  \label{tab:alphabeta}
  \begin{tabular}{ccccc}
    \hline
		Surface & $\alpha$ & $\beta$ & $\sigma_{\rm int}$ & $M_{\rm avg}$\\ 
		\hline
		SALT2-2014 & $0.142 \pm 0.009$ & $2.99 \pm 0.11$ & 0.0979 & 29.979\\ 
		SALT2-2021 & $0.141 \pm 0.009$ & $2.97 \pm 0.11$ & 0.0985 & 29.955\\ 
		\hline
  \end{tabular}
\end{table}

Figure~\ref{fivepanel}d shows $\mu$ before bias corrections are applied. There is a notable shift in $\mu$ with the new surface, where the redshift-dependence is driven primarily by the changes in the recovered supernova colour (Figure~\ref{fivepanel}c). There is a further redshift-dependence in the bias corrections, shown in Figure~\ref{fivepanel}e (discussed further in Appendix~\ref{appendix_bc}). Overall the change in the bias correction is small compared to the changes in colour and peak absolute magnitude. The total effect of the SALT2-2021 surface on the DES-SN3YR distances is shown by the bias-corrected $\mu$ values in Figure~\ref{fivepanel}f.

Using \verb|CosmoMC| results for both surfaces, we find $\Delta w = 0.015$.

\begin{figure}
  \begin{center}
   \includegraphics[width=\columnwidth]{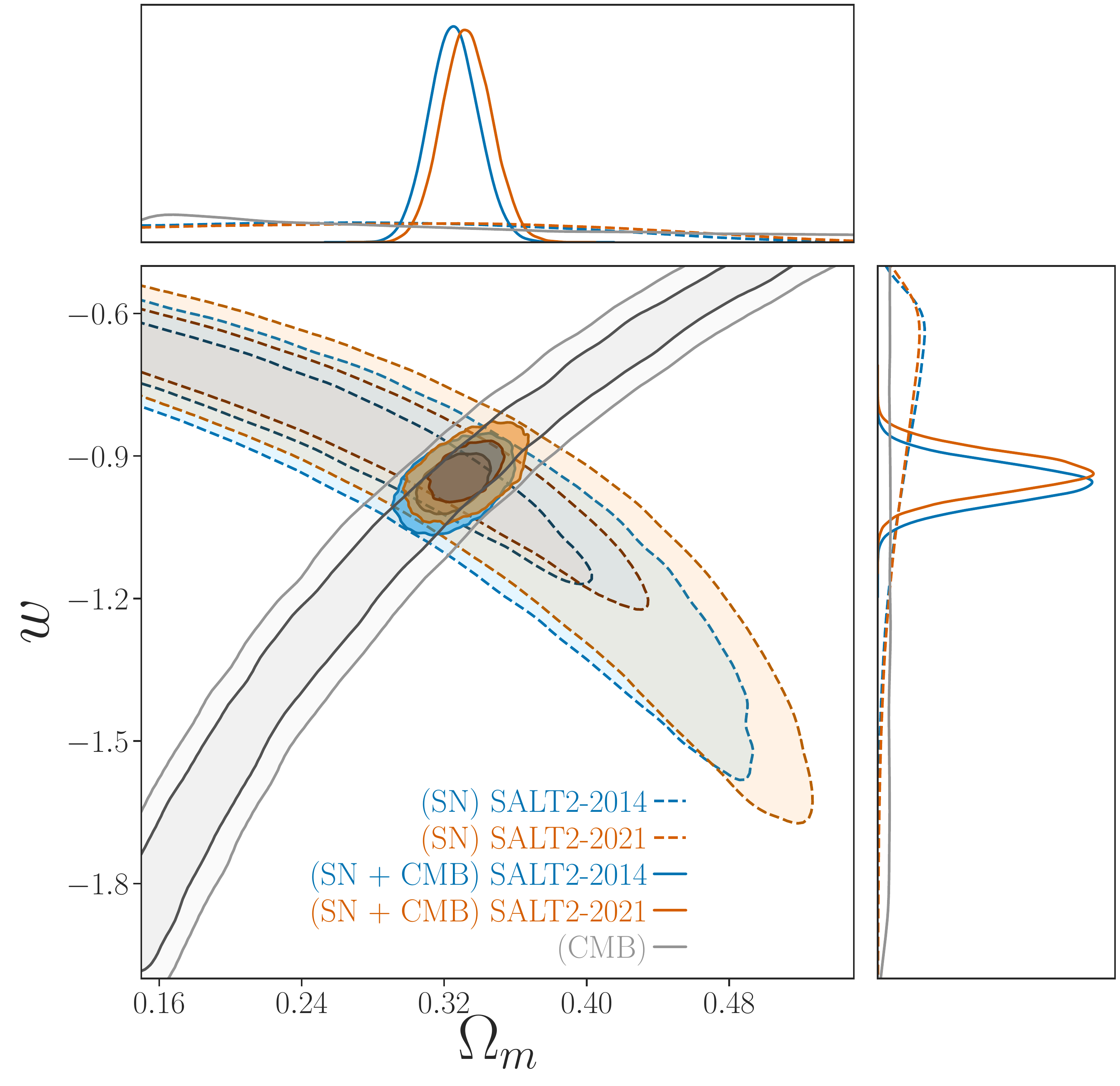}
  \end{center}
  \caption{Contour plot showing the difference in cosmology between the old \textit{(blue)} and new \textit{(orange)} SALT2 surfaces. The individual contours for the DES-SN3YR data \textit{(blue and orange dashed)} and the CMB data \textit{(grey solid)} are also shown.}
  \label{cosmo_snonly}
\end{figure}

\subsection{Simulating Uncertainty in $w$-bias}

\begin{figure*}
  \begin{center}
   \includegraphics[width=2\columnwidth]{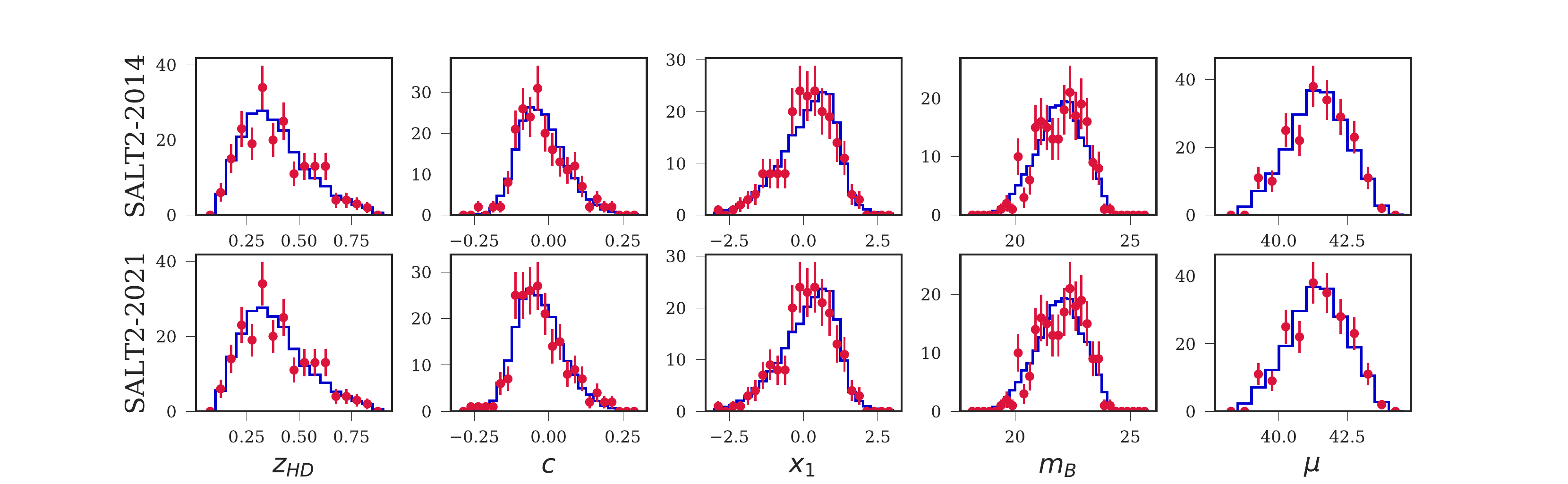}
  \end{center}
  \caption{Parameter distributions for DES-SN data \textit{(red points)} compared with simulated DES SNe Ia \textit{(blue histograms)}. Simulation histograms are scaled to match the size of the data. Error bars show the Poisson error in binned data.}
  \label{des_dist}
\end{figure*}

\begin{figure*}
  \begin{center}
   \includegraphics[width=2\columnwidth]{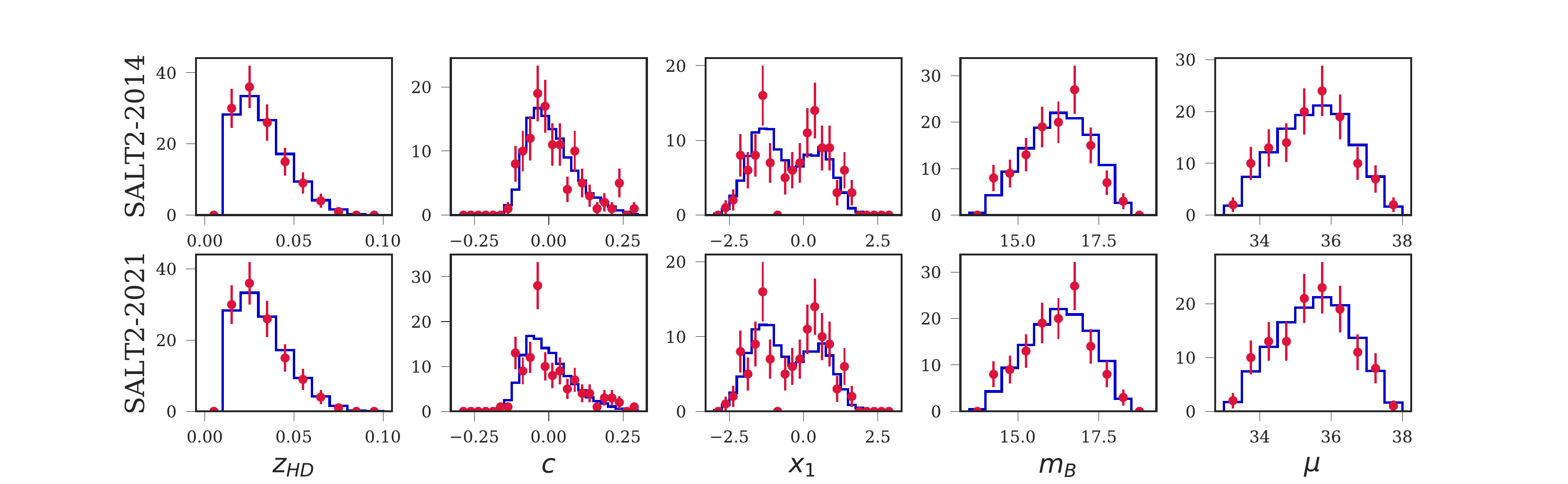}
  \end{center}
  \caption{Parameter distributions for low-$z$ data \textit{(red points)} compared with simulated low-$z$ SNe Ia \textit{(blue histograms)}. Simulation histograms are scaled to match the size of the data. Error bars show the Poisson error in binned data.}
  \label{lowz_dist}
\end{figure*}

To estimate the uncertainty in $\Delta w$ resulting from the change in SALT2 surface, we re-run the analysis on 50 realistic DES-SN3YR-like simulations. Comparisons of various parameter distributions between the data and the simulations are shown in Figure~\ref{des_dist} and Figure~\ref{lowz_dist}, and agree well. For convenience, we replace \texttt{CosmoMC} with a simpler and faster cosmology fitting program in \texttt{SNANA} (\texttt{wfit}). \verb|wfit| uses a $\chi^2$ minimisation to find $w$ and $\Omega_m$ (assuming a $w$CDM model, i.e. a flat universe with a constant $w$ value and cold dark matter). These parameters are constrained with priors based on the cosmic microwave background (CMB) \citep{wmap9}, and fit with only the simulated SN~Ia samples. The $w$ uncertainties produced from \texttt{wfit} are
compared to those from \texttt{CosmoMC} (using a Planck prior) using DES-SN3YR data, and match well ($\sigma_{\texttt{wfit}} = 0.046$; $\sigma_{\texttt{CosmoMC}} = 0.037-0.046$). To avoid biases, the computed CMB prior uses the same cosmological parameters as the SN simulations.

From analysing these 50 simulations, we obtain an RMS (for $\Delta w$) of 0.004. The $\Delta w$ values from the simulations are broadly consistent with that recovered from the DES-SN3YR data.

\section{Discussion and Conclusions}
\label{discussion}

In this paper, we have demonstrated the impact of a revised SALT2 surface on the DES-SN3YR sample. While the overall impact of SALT2-2021 will depend on the specific data sets and analysis methods used, we see clear impacts from the revised surface for the DES-SN3YR sample.

By reducing the Milky Way extinction by 14\%, the SNe in the revised \textit{training} sample have redder intrinsic colours and are fainter, since the corrections are smaller. SALT2-2021 sees a slight change in the average SED, most noticeably at the blue and red ends of the wavelength range (Figure~\ref{SED}), though these regions are not well-constrained in the SALT2 framework with the current training set (Figure~\ref{spec_coverage}). The colour law changes most significantly at the blue end (Figure~\ref{CL}). The mean B-V colour at B band maximum is set to be zero for the SNe used in the SALT2 training \citep{g10_scatter}. Hence, the colour of a SN will depend on the training sample used to derive a SALT2 surface. This is the origin for the offset seen in Figure~\ref{fivepanel}c. On average, the positive residuals indicate that SALT2-2021 lowers the value of $c$ (colour) for SNe~Ia in the data sample (i.e. the SNe population becomes bluer). 

This effect is largely compensated for by the change in absolute magnitude with the new surface (i.e. the SNe population becomes brighter). The impact of the new surface on the other light curve parameters, $x_1$ (stretch) and $m_B$, is small.

These three parameters, along with bias corrections, all contribute to the overall effect on the recovered distance moduli. The change in distance moduli displays a clear redshift dependence, with higher redshift SNe being more affected by the change in SALT2 surface. This is driven mainly by the redshift dependence in the change in supernova colour. A secondary contribution to the redshift dependence in $\mu$ arises from the bias corrections, which is discussed further in Appendix~\ref{appendix_bc}. The change in the measured distance moduli gives $\Delta w = 0.015 \pm 0.004$ with the SALT2-2021 surface. The impact for the DES-SN3YR sample is comparable to many of the systematic uncertainties quoted in the DES-SN3YR analysis \citealt{des3yr_systematics}. The impact of SALT2-2021 on other SNe analyses depends on the sample used and the bias corrections applied. 

We have produced a revised surface for the SN~Ia model SALT2, which should replace the previous surface for future SNe~Ia light curve fitting. In addition, we provide a suite of linearly perturbed SALT2 surfaces that can be used to estimate the effects of calibration uncertainty in the new model. Further improvements with more complete and high-quality data sets are still possible --- in particular, increasing the training sample to include more SNe, particularly in the poorly-sampled UV region. Training on the most extensive, high-quality sample available could reduce statistical uncertainties in the SALT2 model. Additionally, removing SNe that are not measured on their natural systems (that is, the system in which they were observed) could further reduce systematic uncertainties in the model. The uncertainties in the SALT2 model have a significant enough effect to be a critical aspect for future SN studies, particularly if they are trying to constrain a time-variable $w$. In future work, we will examine the systematic and statistical uncertainties of SALT2-2021, and examine how these uncertainties can be reduced by incorporating more modern SNe Ia data sets. 

\section*{Data Availability}
The surface described in this article is publicly available at 10.5281/zenodo.4646495. The data underlying this article are available at \url{https://des.ncsa.illinois.edu/releases/sn}.

\section*{Acknowledgements}

We thank Marc Betoule and the SNLS team for providing us with a copy of the snpca code and their guidance in using it. GT thanks Ryan Ridden-Harper, Philip Wiseman and Tamara Davis for their helpful discussion and comments. This research was supported by an Australian Government Research Training Program (RTP) Scholarship. This project used public archival data from the Dark Energy Survey (DES). Funding for the DES Projects has been provided by the U.S. Department of Energy, the U.S. National Science Foundation, the Ministry of Science and Education of Spain, the Science and Technology Facilities Council of the United Kingdom, the Higher Education Funding Council for England, the National Center for Supercomputing Applications at the University of Illinois at Urbana-Champaign, the Kavli Institute of Cosmological Physics at the University of Chicago, the Center for Cosmology and Astro-Particle Physics at the Ohio State University, the Mitchell Institute for Fundamental Physics and Astronomy at Texas A\&M University, Financiadora de Estudos e Projetos, Funda{\c c}{\~a}o Carlos Chagas Filho de Amparo {\`a} Pesquisa do Estado do Rio de Janeiro, Conselho Nacional de Desenvolvimento Cient{\'i}fico e Tecnol{\'o}gico and the Minist{\'e}rio da Ci{\^e}ncia, Tecnologia e Inova{\c c}{\~a}o, the Deutsche Forschungsgemeinschaft, and the Collaborating Institutions in the Dark Energy Survey.
The Collaborating Institutions are Argonne National Laboratory, the University of California at Santa Cruz, the University of Cambridge, Centro de Investigaciones Energ{\'e}ticas, Medioambientales y Tecnol{\'o}gicas-Madrid, the University of Chicago, University College London, the DES-Brazil Consortium, the University of Edinburgh, the Eidgen{\"o}ssische Technische Hochschule (ETH) Z{\"u}rich, Fermi National Accelerator Laboratory, the University of Illinois at Urbana-Champaign, the Institut de Ci{\`e}ncies de l'Espai (IEEC/CSIC), the Institut de F{\'i}sica d'Altes Energies, Lawrence Berkeley National Laboratory, the Ludwig-Maximilians Universit{\"a}t M{\"u}nchen and the associated Excellence Cluster Universe, the University of Michigan, the National Optical Astronomy Observatory, the University of Nottingham, The Ohio State University, the OzDES Membership Consortium, the University of Pennsylvania, the University of Portsmouth, SLAC National Accelerator Laboratory, Stanford University, the University of Sussex, and Texas A\&M University. 
Based in part on observations at Cerro Tololo Inter-American Observatory, National Optical Astronomy Observatory, which is operated by the Association of Universities for Research in Astronomy (AURA) under a cooperative agreement with the National Science Foundation.
This work was completed in part with resources provided by the University of Chicago Research Computing Center.
DB acknowledges support for this work was provided by NASA through the NASA Hubble Fellowship grant HST-HF2-51430.001 awarded by the Space Telescope Science Institute, which is operated by Association of Universities for Research in Astronomy, Inc., for NASA, under contract NAS5-26555.
Figure~\ref{cosmo_snonly} was produced with \texttt{ChainConsumer} \citep{Hinton2016}.




\bibliographystyle{mnras}
\bibliography{refs} 



\appendix

\section{Process Verification}
\label{verification}

\begin{figure}
  \begin{center}
   \includegraphics[width=\columnwidth]{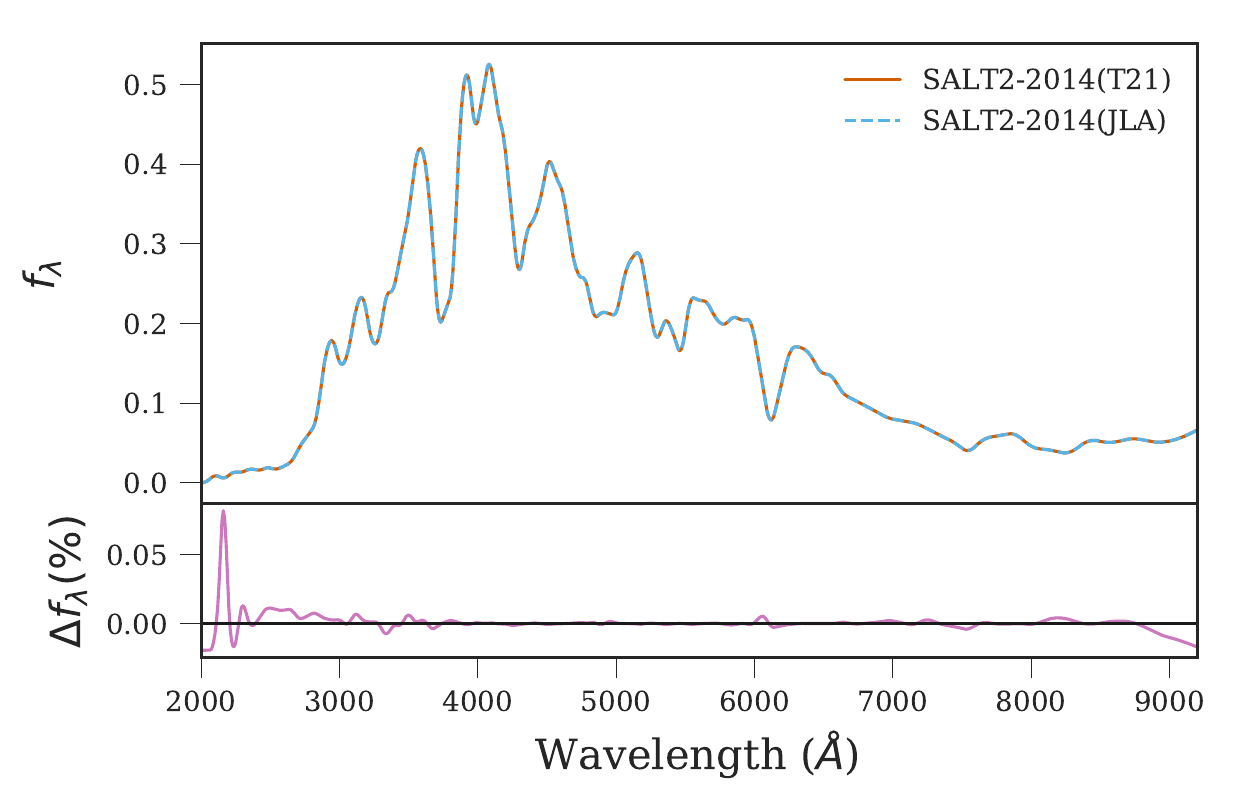}
  \end{center}
  \caption{The trained spectral energy distribution of a SN~Ia ($c$=0, $x_1$=0) for the SALT2-2014(JLA) surface was reproduced with the accuracy shown above (the two lines are directly overlaid). Our ability to reproduce this surface gives us confidence in the training process.}
  \label{SED_JLA}
\end{figure}

As a check on our training process, we reproduce the most recent surface, SALT2-2014(JLA), using \verb|snpca|. This is shown in Figure~\ref{SED_JLA}, where our reproduced copy of the surface is denoted SALT2-2014(T21). 

\begin{figure*}
  \begin{center}
   \includegraphics[width=2\columnwidth]{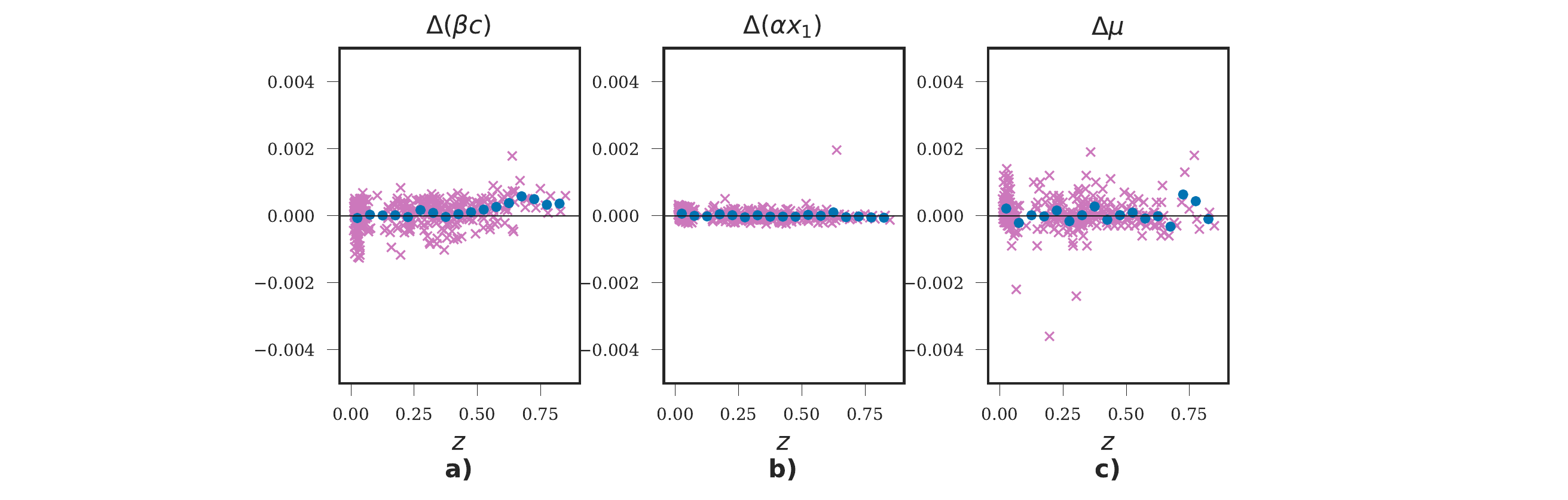}
  \end{center}
  \caption{Differences in fitted parameters between the published JLA surface and the JLA surface we trained, plotted versus redshift. Differences are given as e.g. $\Delta \mu = \mu_{2014(\rm T21)} - \mu_{2014(\rm JLA)}$.
  Parameters are \textit(from left to right): $\beta~\times$~colour, $\alpha~\times$~stretch, and $\mu$ after bias corrections. Each SN in the fitted sample is plotted as a pink cross, with the binned averages plotted as blue circles \textit{(nbins=20)}. One $\Delta \mu$ outlier is not shown.}
  \label{threepanel}
\end{figure*}

As a further check, we refit the SN sample in the DES-SN3YR data release\footnote{Available from \url{https://des.ncsa.illinois.edu/releases/sn}} \citep{deslc} using the same SALT2-2014(JLA) version that they used.\footnote{Available from \url{http://supernovae.in2p3.fr/salt/}} Differences in the light curve parameters (and resulting SN distances) between our results and the published results are at the millimag level and are shown in Figure~\ref{threepanel}. We have not tracked down the reasons for the differences, but they are small enough to have a negligible impact on the derived distances --- note the scale on the axes are an order of magnitude smaller than those from the change in surface in Figure~\ref{fivepanel}. We therefore have confidence that the training and fitting methods used for the SALT2-2021 surface are consistent with the methods used in the JLA, and any differences from SALT2-2021 are from the revised input data and not from our implementation of the process.

\section{Analysing Bias Corrections}
\label{appendix_bc}

The difference in the recovered (bias-corrected) distance moduli for SALT2-2021 versus SALT2-2014 displays a noticeable redshift dependence, where higher-redshift SNe Ia are more affected by the change in surface (Figure~\ref{fivepanel}f). Some of this redshift dependence comes from the bias corrections applied ($\Delta \delta \mu_{\rm bias}$, Figure~\ref{fivepanel}e). Here, we briefly examine the reasons for this effect.

\begin{figure*}
  \begin{center}
   \includegraphics[width=2\columnwidth]{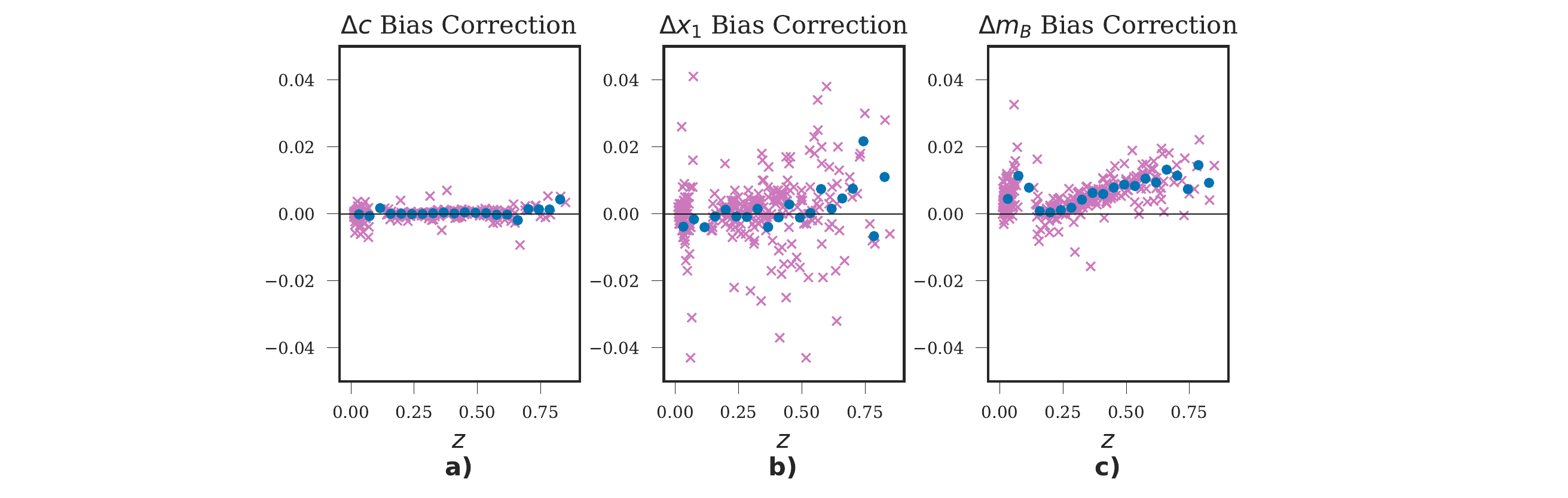}
  \end{center}
  \caption{Differences in bias corrections of fitted parameters between the old and new SALT2 surfaces, plotted versus redshift. Differences are given as e.g. $\Delta c = c_{2021} - c_{2014}$. Parameters are \textit(from left to right): colour bias corrections, stretch bias corrections, and $m_B$ bias corrections. Each SN in the fitted sample is plotted as a pink cross, with the binned averages plotted as blue circles \textit{(nbins=20)}.}
  \label{threepanel_bc}
\end{figure*}

Figure~\ref{threepanel_bc} shows the difference in the bias corrections applied for parameters fitted using SALT2-2021 versus SALT2-2014. The $mB$ parameter bias correction displays the strongest redshift-dependent trend, and looks to be the main contributor to the effect that we see in Figure~\ref{fivepanel}e. Interestingly, there is no obvious redshift-dependent trend in the $c$ bias corrections, though the fitted colour parameters from the DES-SN3YR data show a clear redshift dependence (Figure~\ref{fivepanel}c). We also note that the redshift-dependent shift in bias corrections persists when performing 1D (rather than 5D) bias corrections on the DES-SN3YR sample.

We investigate the origins of this redshift dependent change in bias corrections (caused by the change in SALT2 surface from SALT2-2014 to SALT2-2021) by looking at the bias corrections in the DES-SN3YR data from other SALT2 surfaces. We create ``hybrid'' surfaces that combine components ($C_L$, $M_0$, $M_1$) from SALT2-2014 and SALT2-2021, to see if a particular component is responsible for the effect. While the colour law appears to have the strongest effect on the bias corrections, it does not fully explain the shift in bias corrections, which is only seen when using all three SALT2-2021 components in a surface. We also examine the bias corrections when fitting over a restricted wavelength range (4000-7000\AA), and rule out fitting data in the UV range as the cause of the redshift dependent $\Delta \delta \mu_{\rm bias}$. Either the three SALT2-2021 components have some extra effect when applied in combination, or there is some other, unknown factor that is affecting the bias corrections.

\bsp	
\label{lastpage}
\end{document}